\documentclass[pra,twocolumn,twoside,showpacs,preprintnumbers,amsfonts,amsmath,amssymb,floatfix]{revtex4}

\usepackage{graphicx}
\usepackage{multirow}
\usepackage{color}
\usepackage{dcolumn}

\newcommand{\bra}[1]{\ensuremath{\left<#1\right|}}
\newcommand{\ket}[1]{\ensuremath{\left|#1\right>}}
\newcommand{\qmop}[1]{\ensuremath{\mathbf#1}}

\newcommand{\hho}{H$_2$O}
\newcommand{\ddo}{D$_2$O}
\newcommand{\hdo}{HDO}

\newcommand{\minitab}[2][l]{\begin{tabular}{#1}#2\end{tabular}}

\makeatletter 
\newcommand{\roem}[1]{
        \textrm{
                \@roman{#1}
        }\ 
        \ignorespaces 
}
\newcommand{\Roem}[1]{
        \textrm{
                \@Roman{#1}
        }\ 
        \ignorespaces 
}
\renewcommand{\Roem}[1]{\textrm{\@Roman{#1}}}
\makeatother 

\begin{document}
\title{Cold guided beams of water isotopologs}
\author{M. Motsch}
\author{L.D. van Buuren}
\author{C. Sommer}
\author{M. Zeppenfeld}
\author{G. Rempe}
\author{P.W.H. Pinkse}
\email{pepijn.pinkse@mpq.mpg.de}
\affiliation{Max-Planck-Institut f{\"u}r Quantenoptik, Hans-Kopfermann-Stra{\ss}e 1, 85748 Garching, Germany}



\begin{abstract}
Electrostatic velocity filtering and guiding is an established technique to produce high fluxes of cold polar molecules. In this paper we clarify different aspects of this technique by comparing experiments to detailed calculations. In the experiment, we produce cold guided beams of the three water isotopologs H$_2$O, D$_2$O and HDO\@. Their different rotational constants and orientations of electric dipole moments lead to remarkably different Stark shift properties, despite the molecules being very similar in a chemical sense. Therefore, the signals of the guided water isotopologs differ on an absolute scale and also exhibit characteristic electrode voltage dependencies. We find excellent agreement between the relative guided fractions and voltage dependencies of the investigated isotopologs and predictions made by our theoretical model of electrostatic velocity filtering.
\end{abstract}

\maketitle

\section{Introduction and Motivation}
\label{sec:intro}
Cold polar molecules offer fascinating perspectives for research, e.g.\ in cold and ultracold chemistry, precision measurements for tests of fundamental symmetries, and quantum information (see, e.g., the special issue on cold polar molecules \cite{Dulieu2006}). Since molecules are in general inaccessible to direct laser cooling, new methods are needed. Indirect methods such as association of molecules from ultracold atomic ensembles by photoassociation \cite{Schloeder2001,Kerman2004,Jones2006} or using magnetic Feshbach resonances \cite{Inouye2004,Stan2004,Kohler2006} have the advantage of producing polar molecules directly at ultralow translational temperatures. Although the molecules are produced in highly excited vibrational states, it has been shown that it is possible to transfer them down the vibrational ladder to create stable ultracold molecules \cite{Ospelkaus2008,Danzl2008,Deiglmayr2008,Lang2008}. However, these techniques are limited to a few species which can be forged together from laser-coolable atoms. Direct methods, which can be applied to naturally occurring polar molecules, include buffer-gas cooling \cite{Weinstein1998}, electric \cite{Bethlem1999,Bethlem2002,vandeMeerakker2008} and optical \cite{Fulton2004,Fulton2006} Stark deceleration, magnetic deceleration \cite{Vanhaecke2007,Narevicius2008}, collisions with counterpropagating moving surfaces \cite{Narevicius2007} or collision partners \cite{Elioff2003}, rotating nozzles \cite{Gupta1999}, velocity filtering of large molecules by rotating mechanical filters \cite{Deachapunya2008}, and velocity filtering by an electrostatic quadrupole guide \cite{Rangwala2003,Junglen2004a}.

Electrostatic velocity filtering offers the advantage of being a simple technique delivering continuous beams of high flux; guided beams of more than 10$^{10}$ molecules/s have been produced for ND$_3$ and H$_2$CO \cite{Rangwala2003,Junglen2004a}. This electrostatic guiding and filtering technique is well suitable for collision experiments \cite{Willitsch2008}, since it produces a high continuous flux with only few rotational states contributing \cite{Rieger2006,Motsch2007}. It has been combined in a natural way with buffer-gas cooling to increase the purity of the guided beam \cite{vanBuuren2008}. Furthermore, it is applicable to many molecular species, as long as they have populated low-field-seeking (lfs) states in the thermal source. For high-field-seeking (hfs) states, velocity filtering in the guide is also possible by changing from a static quadrupole field to a time-dependent dipole field \cite{Junglen2004}.

The filtering by an electric guide strongly depends on the Stark shift properties of the molecules used. In general, dissimilar molecules not only have different Stark shifts, but also different physical and chemical properties, which play an important role in the source and in the detection process. Therefore, it is difficult in general to compare such measurements. In this paper we present experiments performed with the three water isotopologs {\hho}, {\ddo} and {\hdo}\@. These molecules have comparable masses and chemical properties but differ in their rotational constants and dipole moments. This allows the study of the velocity filtering process without large uncontrollable systematic effects. Although these isotopologs seem very similar, they show surprisingly different behavior when exposed to external electric fields. Their different Stark shift properties are experimentally revealed as characteristic dependencies of their detector signals and velocity distributions on the applied electrode voltages. The experimental work is accompanied by a theoretical description of the filtering process. We find a good agreement between the predictions of our calculations and the experiments, concluding that the filtering process is well described by the presented model.

The paper is organized as follows: In Section \ref{sec:velfilter} we review the filtering process in an electric guide for polar molecules. In Section \ref{sec:calcStark} we present the theory for Stark shifts of asymmetric rotor molecules, which is applied to the water isotopologs {\hho}, {\ddo} and {\hdo}\@. In Section \ref{sec:calcGuide} it is shown how the calculated Stark shifts of {\hho}, {\ddo} and {\hdo} and the described filtering properties of the guide can be used to make predictions of the guided flux. The experimental setup used for the guiding experiments is described in Section \ref{sec:exp}. From detailed measurements presented in Section \ref{sec:results} we determine relative detector signals and velocity distributions of cold and slow water molecules.

\section{Velocity filtering}
\label{sec:velfilter}

To calculate the relative guided fluxes of the different water isotopologs, it is necessary to review some basics of velocity filtering by an electric guide. For guiding of low-field-seeking molecules, the guide electrodes are charged to positive and negative high voltages in a quadrupolar configuration (as shown in  Fig.~\ref{fig:ExpSetup} in Section \ref{sec:exp}). This creates an electric field minimum in the center, surrounded by a linearly increasing electric field, up to a certain maximum trapping field. In this electric field minimum molecules in lfs states can be trapped in transverse direction and guided.

Molecules from a thermal reservoir at temperature $T$ are injected into the guide. Their velocities are described by a three-dimensional Maxwell-Boltzmann velocity distribution \begin{subequations}
\begin{equation}
f(v)dv = \frac{4}{\sqrt{\pi}\alpha^3} v^2 \exp(-v^2/\alpha^2) dv
\end{equation}
and by a one-dimensional velocity distribution
\begin{equation}
f(v_{x,y,z})dv_{x,y,z} = \frac{1}{\sqrt{\pi}\alpha} \exp(-v_{x,y,z}^2/\alpha^2) dv_{x,y,z}
\end{equation}
\end{subequations}
with most probable velocity $\alpha = \sqrt{2k_BT/m}$, velocity components $v_i\;(i=x,y,z)$ and total velocity $v=\sqrt{v_x^2+v_y^2+v_z^2}$. The velocity distribution of molecules coming from the nozzle is given by
\begin{equation}
P(v_z)dv_z =  \frac{2}{\alpha^2} v_z \exp(-v_z^2/\alpha^2) dv_z,
\end{equation}
where the $z$-direction is defined to be oriented along the guide. The factor $v_z$ enters since one is now considering the velocity distribution in the flux out of the nozzle and not in a fixed volume.

The guided fraction of molecules can be calculated as the part of molecules injected into the guide with transverse and longitudinal velocities below certain transverse and longitudinal cut-off velocities, which depend on the Stark shift of the molecules and the properties of the guide. For a molecule with a given Stark shift $\Delta W^s(E_{max})$ at the maximum trapping field $E_{max}$, a maximum transverse velocity $v_{max}=\sqrt{2\Delta W^s/m}$ exists. If the transverse velocity $v_\bot$ of the molecule exceeds $v_{max}$, it is lost from the guide. In the following it is assumed that the molecule is injected in the center of the guide. If molecules enter the guide off-center they have acquired already some potential energy which reduces the maximum trappable transverse velocity. Furthermore we assume mixing of the transverse degrees of freedom, whereas longitudinal and transverse degrees of freedom are not coupled. In a full numerical analysis this assumption is dropped and yields only small modifications to the guided flux. Filtering on longitudinal velocity $v_l$ is realized by bending the guide. In the bend, molecules with a velocity below the maximum longitudinal guidable velocity $v_{l,max}$ remain trapped. The longitudinal and radial maximal velocities are connected by the centrifugal force acting on the molecules in the bend, leading to $v_{l,max} \propto v_{max}$. More specific, the guide has a free inner radius $r$, at which the maximum trapping field is reached. Since the trapping potential increases linearly in radial direction, the maximum restoring force acting on a molecule is given by $F_{r,max} = \Delta W^s(E_{max})/r$. The maximum longitudinal velocity is then obtained by equating the centrifugal force in the bend of radius $R$ and the restoring force, resulting in $v_{l,max} = \sqrt{\Delta W^s(E_{max}) R/r\,m} = \sqrt{R/2r}\,v_{max}$. Note that for increasing longitudinal velocity the maximum trappable transverse velocity decreases. As basically every particle is guided if $v_l < v_{l,max}$ and $v_\bot < v_{max}$, this results in higher efficiencies as compared to filtering by, e.g., rotating filter wheels and apertures.

To calculate the guided flux, the velocity distributions of molecules injected into the guide are integrated to the maximum trappable velocity. When performing the integration in the limit of small cut-off velocities $v_{max}$ and $v_{l,max}$ compared to the thermal velocity $\alpha$, the exponential $\exp(-v^2/\alpha^2)$ in the thermal velocity distributions can be replaced by 1. The guided flux $\Phi$ of a molecular state with a Stark energy $\Delta W^s$ is then given by
\begin{multline}
\label{eq:VelFilteringFlux}
\Phi =\\\int\limits_{v_x=-v_{max}}^{v_{max}} \int\limits_{v_y=-v_{max}}^{v_{max}} \int\limits_{v_z=0}^{v_{l,max}} f(v_x) f(v_y) P(v_z) dv_x dv_y dv_z\\
\propto v_{max}^4 \propto (\Delta W^s)^2,
\end{multline}
where we have neglected the dependence of the transverse cut-off velocity on the longitudinal velocity, and $\Phi$ is normalized to the flux out of the nozzle. The guided flux can hence be described by a function $f$ which gives the fraction of guidable molecules for a given electric field, $\Phi=f(\Delta W^s) \propto (\Delta W^s)^2$. The dependence of the guidable fraction given by $f(\Delta W^s)$ and therefore the dependence of the guided flux on the molecular Stark shift allows us to infer characteristic Stark shift properties from measurements of the guided flux. When varying the electrode voltage and hence the guiding electric field, the flux will change depending on the molecules' Stark shift. As will be shown in Section~\ref{sec:calcStark}, molecules can exhibit linear or quadratic Stark shifts, depending on the molecular properties. In Section~\ref{sec:calcGuide} the influence of these Stark shift behavior on the guided flux will be discussed in detail. In short, the flux scales quadratically with electrode voltage for molecules with a linear Stark effect and quartic for molecules with a quadratic Stark effect, as can be seen from Eq.~(\ref{eq:VelFilteringFlux}) and as already described in \cite{Rangwala2003,Junglen2004a,Rieger2006}.

\section{Stark shifts}
\label{sec:calcStark}

As shown in Section \ref{sec:velfilter}, the guided flux of molecules in a certain rotational state is determined by the Stark shift of that state. The total flux is then given by the sum over the contributions of all states thermally populated in the source. To make a theoretical prediction for guided fluxes of the different water isotopologs {\hho}, {\ddo} and {\hdo}, their Stark shifts are calculated by numerical diagonalization of the asymmetric rotor Hamiltonian in the presence of an external electric field, following the procedure given by Hain \emph{et al.} \cite{Hain1999}. For completeness, the main steps of the calculation are summarized in Section~\ref{subsec:calcStark:calc}. A detailed understanding of this theoretical part of the paper is not crucial for the discussions following. Section~\ref{subsec:calcStark:Applications} shows how the molecular properties manifest themselves in the Stark shifts, and illustrates this with the example of the different water isotopologs.

\subsection{Calculation of Stark Shifts}
\label{subsec:calcStark:calc}
\begin{figure}
\includegraphics[]{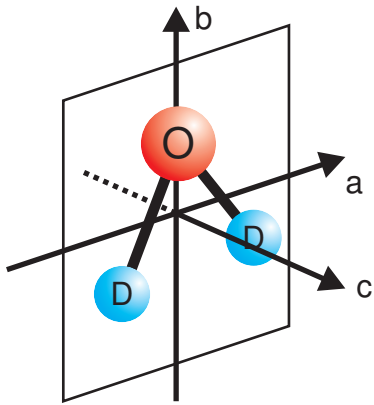}
\caption{\label{fig:molecule}(Color online). Orientation of the rotational axes in {\ddo}. The electric dipole moment is oriented along the $b$-axis.}
\end{figure}

The molecular properties relevant for the calculations are listed in Table \ref{tab:rotConst}. $A$, $B$ and $C$ are the rotational constants along the principal axes, where $A$ ($C$) is oriented along the axis with largest (smallest) rotational constant, $A \geq B \geq C$. The orientation of the rotational axes in {\ddo} is illustrated in Fig.~\ref{fig:molecule}.

\begin{table}
\begin{ruledtabular}
\begin{tabular}{lc|ddd}
                &       &
\multicolumn{1}{c}{\hho}   &
\multicolumn{1}{c}{\hdo}   &
\multicolumn{1}{c}{\ddo}   \\
\hline
\multirow{3}{*}{\minitab[l]{Rotational\\constants\\ $[\mathrm{cm}^{-1}]$}}&
$A$         &27.79    &23.48     &15.39       \\
&$B$        &14.50    &9.13      &7.26      \\
&$C$        &9.96     &6.40      &4.85      \\
\hline
\multirow{3}{*}{\minitab[l]{Dipole moment\\components \\ $[\mathrm{Db}]$}}&
$\mu_A$     &0        &0.66     &0       \\
&$\mu_B$    &1.94     &1.73      &1.87       \\
&$\mu_C$    &0        &0         &0      \\
\end{tabular}
\end{ruledtabular}
\caption{\label{tab:rotConst}Rotational constants and components of the electric dipole moment along the principal axes for the different water isotopologs {\hho}, {\ddo} and {\hdo} \cite{Townes:Microwave,Brittain1972}.}
\end{table}

The molecular Hamiltonian in the absence of external electric fields can be written as
\begin{subequations}
\begin{equation}
\label{eq:AsymmRotHamiltonian}
H_{rot}=\frac{1}{2}(A+C)\qmop{J^2}+\frac{1}{2}(A-C)H(\kappa),
\end{equation}
with the reduced Hamiltonian
\begin{equation}
\label{eq:AsymmRotRedHamiltonian}
H(\kappa)=\qmop{J_a^2}+\kappa \qmop{J_b^2}-\qmop{J_c^2}.
\end{equation}
\end{subequations}
The constant $\kappa = (2B-A-C)/(A-C)$ is the so called asymmetry parameter, taking on the value $-1$ in the limit of the prolate symmetric top and $+1$ in the limit of the oblate symmetric top. For calculations this Hamiltonian is expressed in the symmetric rotor basis. The matrix elements of the reduced Hamiltonian $H(\kappa)$ [Eq.~(\ref{eq:AsymmRotRedHamiltonian})] in the symmetric top basis $\left\{\ket{JKM}\right\}$ are given by
\begin{subequations}
\label{eq:AsymmRotRedHamiltonianMatrixElements}
\begin{gather}
\bra{JKM}H(\kappa)\ket{JKM} = F \left[ J(J+1)-K^2 \right] +GK^2,\\
\bra{J,K\pm2,M}H(\kappa)\ket{JKM} = H \left[ f(J,K\pm1) \right]^{1/2},
\end{gather}
with
\begin{multline}
f(J,K\pm1)= \frac{1}{4} \left[ J(J+1)-K(K\pm1)\right]\\
            \times\left[ J(J+1)-(K\pm1)(K\pm2)\right]
\end{multline}
\end{subequations}
and $F,G,H$ supplied in Table~\ref{tab:Representations} and \ref{tab:FGH} in the appendix. Only symmetric rotor states with $\Delta K=0,\pm2$ are coupled by the asymmetric rotor Hamiltonian as is seen from Eqs. (\ref{eq:AsymmRotRedHamiltonianMatrixElements}). The eigenstates $A_{J\tau M}$ and energies $W_{J\tau M}$ of the asymmetric rotor can be found by diagonalization of the full asymmetric rotor Hamiltonian Eq.~(\ref{eq:AsymmRotHamiltonian}), yielding
\begin{equation}
W_{J\tau M} = \frac{1}{2} (A+C) J (J+1) + \frac{1}{2} (A-C) W_{J\tau M}(\kappa)
\end{equation}
and
\begin{equation}
A_{J\tau M} = \sum_K a_K^{J\tau M} \Psi_{JKM}.
\end{equation}
The eigenstates $A_{J\tau M}$ are expressed as linear superposition of symmetric rotor wave functions $\Psi_{JKM}$. Note that the total angular momentum quantum number $J$ and its projection on a space-fixed axis $M$ are still good quantum numbers in the field-free asymmetric rotor, in contrast to $K$. This can already be seen from the asymmetric rotor Hamiltonian which does not depend on $M$ and which commutates with \qmop{J^2}. The pseudo quantum number $\tau=K_{a}-K_{c}$ is used to label the asymmetric rotor states in ascending order in energy. $\tau$ is directly related to the quantum numbers $K_{a}$ and $K_{c}$ in the prolate and oblate limiting case of the symmetric top. From this the symmetry properties of the asymmetric rotor states, represented by the symmetry species of the four group $D_2$ \cite{BunkerMolSymm} (sometimes also referred to as $V(a,b,c)$), can be derived \cite{King1943,Cross1944}.

An external field lifts the degeneracy between the  $M$-sublevels of a state \ket{J,\tau}. Different $J$ states are also coupled now, leaving $M$ as the only good quantum number. In general, an asymmetric top molecule can possess components of its dipole moment $\mu$ along all three principal axes in the body-fixed frame , $\vec{\mu} = \sum_g \mu_g \hat{e}_g,\;(g=a,b,c)$. As can be seen from Table~\ref{tab:rotConst}, {\hho} and {\ddo} have a dipole moment component only along the $b$-axis (see Fig.~\ref{fig:molecule}), whereas in {\hdo} there are components along the $a$- and $b$-axis. This is important for their Stark shift properties, since different components promote couplings between rotational energy levels of different symmetry species. We define the electric field to be directed along the $Z$-axis in the space-fixed frame $(F=X,Y,Z)$, $\vec{E} = E_Z \hat{e}_Z$. Then, the interaction Hamiltonian is given by
\begin{equation}
H_s = E_Z \sum_g \Phi_{Zg} \mu_g,
\end{equation}
where the direction cosines $\Phi_{Fg}$ connect the space-fixed to the molecule-fixed frame. Since the direction cosines are tabulated for symmetric rotor wave functions only (see Table \ref{tab:PhiFg} in Appendix \ref{sec:AppendixTables}; note there are some misprints in the $\Phi_{Fg}$ tabulated in  \cite{Hain1999} which are corrected here.), their matrix elements with respect to the asymmetric rotor states have to be constructed. This can be done using the expansion of the asymmetric rotor wave functions in terms of symmetric rotor states and results in
\begin{multline}
\label{eq:StarkMatrixElements}
\bra{J\tau M} \Phi_{Zg} \ket{J'\tau' M'} = \\
\bra{J}\Phi_{Zg}\ket{J'} \bra{JM}\Phi_{Zg}\ket{J'M'}\delta_{MM'}\\
\times \sum_{KK'}a_K^{J\tau M}a_{K'}^{J'\tau' M'}\bra{JK}\Phi_{Zg}\ket{J'K'}.
\end{multline}
Note that $H_s$ only couples states with $\Delta M=0$, due to the choice of $\vec{E}$ along the $Z$-direction, $\vec{E} = E_Z \hat{e}_Z$. By diagonalization of the total Hamiltonian $H_{rot}+H_s(E_Z)$ the eigenstates and eigenenergies of the asymmetric rotor in the presence of an external field can be calculated. The Stark shift $\Delta W^s(E_Z)$ is then simply given by the difference between the total energy in the external electric field and the zero field energy.

\subsection{Discussion of Stark Shifts}
\label{subsec:calcStark:Applications}

\begin{figure}
\includegraphics[width=1.\columnwidth]{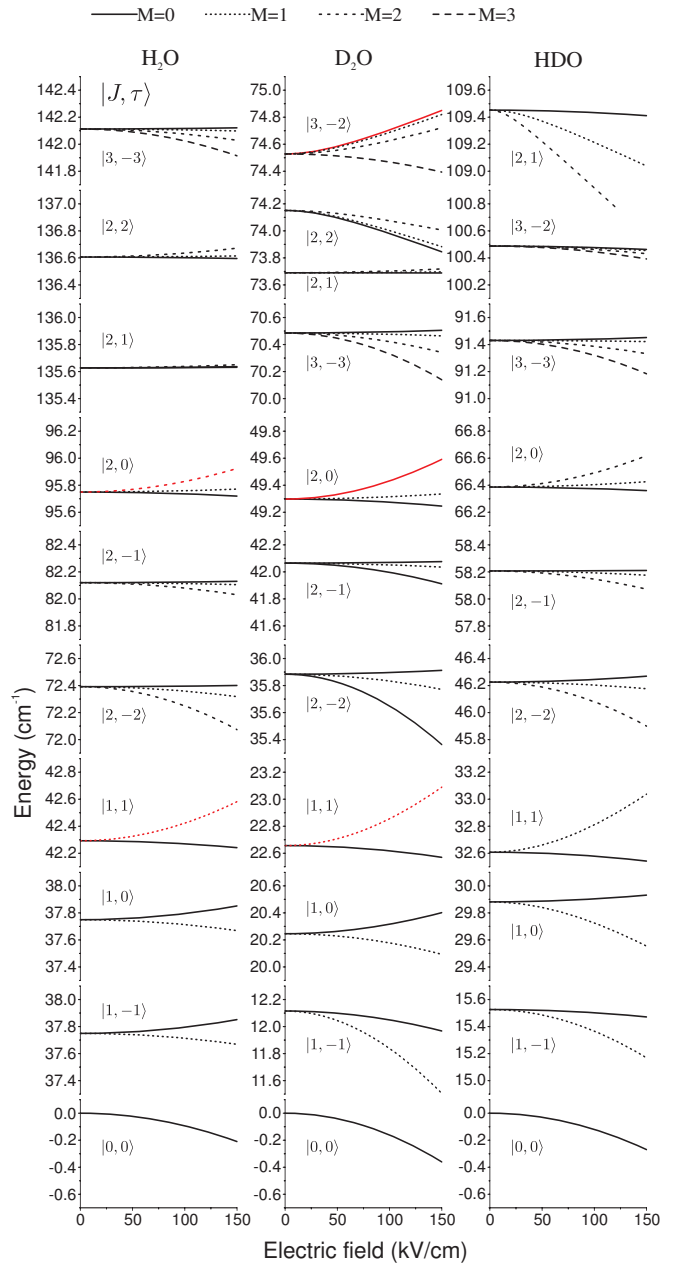}
\caption{\label{fig:StarkShiftCurves}(Color online). Energy of rotational states in an external electric field for the water isotopologs {\hho}, {\ddo} and {\hdo}\@. The same vertical scale is used throughout the figure. States with large contributions in the guided beam are plotted with thick red lines and marked by a star.}
\end{figure}

\begin{figure}
\includegraphics[width=1.\columnwidth]{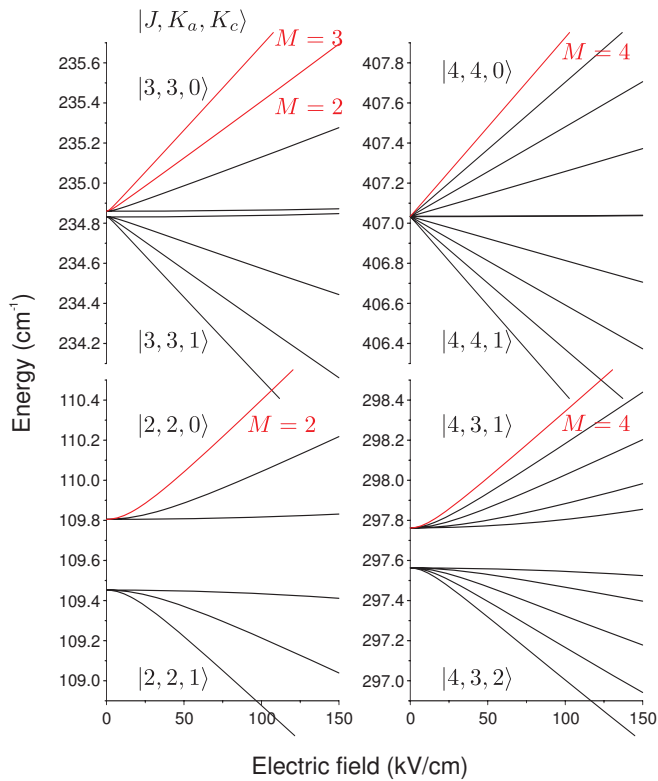}
\caption{\label{fig:StarkShiftCurvesHDOlin}(Color online). Lowest energy rotational states with linear Stark shifts of {\hdo}\@. The same vertical scale is used throughout the figure. States with large contributions in the guided beam are plotted with thick red lines and marked by a star. In this figure quantum numbers $K_a$ and $K_c$ are used to indicate the close connection to the prolate symmetric rotor where these states are degenerate, giving rise to linear Stark shifts.}
\end{figure}

\begin{figure}
\includegraphics[width=1\columnwidth]{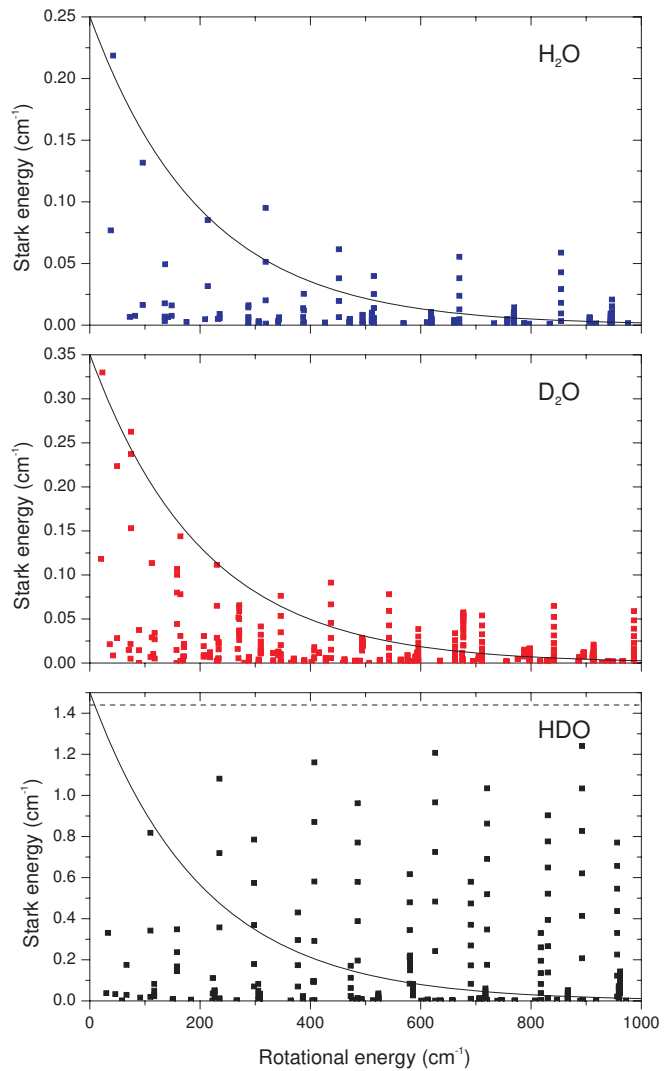}
\caption{\label{fig:StarkShifts}(Color online). Stark shifts of the different water isotopologs {\hho}, {\ddo} and {\hdo} at an electric field of 130\,kV/cm. The solid line indicates the Boltzmann factor for a source temperature of 293\,K. Only lfs states are shown. Note the different vertical scales. The different $M$ states of a rotational state \ket{J,\tau,M} all have the same rotational energy but different Stark shifts, thus forming a vertical sequence. Note that in the case of {\hdo} the Stark shifts approach a saturation value indicated by the dashed line for increasing rotational energy, corresponding to the full alignment of the dipole moment component along the $a$-axis on the electric field axis. In the case of the $b$-type rotors {\hho} and {\ddo} the Stark shifts decrease with increasing rotational energy.}
\end{figure}

The energies of the rotational states of the different water isotopologs in an external electric field, calculated by the procedure described in Section~\ref{subsec:calcStark:calc}, are shown in Fig.~\ref{fig:StarkShiftCurves} and \ref{fig:StarkShiftCurvesHDOlin}. Several features are directly evident when comparing the different isotopologs. First of all, the calculations predict quadratic Stark shifts for {\hho} and {\ddo}, while for {\hdo} linear Stark shifts are found as well. This is caused by the different orientations of the electric dipole moments in the molecules with respect to the main axes of the molecule. In {\hho} and {\ddo} the dipole moment is only oriented along the $b$-axis. In {\hdo} there is additionally a component along the $a$-axis. This is important since in a more prolate asymmetric rotor, as is the case for the different water isotopologs ({\hho} $\kappa=-0.49$, {\ddo} $\kappa=-0.54$, {\hdo} $\kappa=-0.68$), the states with the same $K_a$ quantum number are near degenerate for increasing $J$ and $K_a$ quantum numbers. It is exactly these states which are coupled by a dipole moment along the $a$-axis in an external electric field. Hence, once the coupling between these states due to the electric field becomes larger than their asymmetric rotor splitting the Stark shift becomes linear. This behavior is illustrated in Fig. \ref{fig:StarkShiftCurvesHDOlin} for the lowest energy rotational states of {\hdo} exhibiting a mainly linear Stark shift. For {\hho} and {\ddo} the situation is completely different, here the Stark shifts stay quadratic and even become smaller with increasing rotational energy as can be seen from Fig.~\ref{fig:StarkShifts}.

A second observation which can be made from Figs.~\ref{fig:StarkShiftCurves}--\ref{fig:StarkShifts} is that the magnitude of the Stark shifts of the isotopologs also differ. {\hdo} exhibits the largest Stark shifts, as is to be expected because of their linear character. Besides, the Stark shifts of {\hho} are significantly smaller than those of {\ddo}, although both molecules have similar size dipole moments and hence similar couplings between rotational states. The reason for this is also found in the rotational constants. Since {\ddo} has smaller rotational constants than {\hho}, the energy levels are closer together, leading to larger Stark shifts for comparable couplings, as can be seen from perturbation theory.

The third observable feature is  that the Stark shifts also show a different behavior with increasing rotational energy for the different isotopologs. In {\hdo}, the states \ket{J,\tau,M}=\ket{J,J,J} exhibit the maximum Stark shifts which approach a constant value given by the maximum possible projection of the dipole moment component $\mu_a$ along the $a$-axis on the electric field axis as indicated in Fig.~\ref{fig:StarkShifts}. The Stark shifts in {\hho} and {\ddo} decrease with increasing rotational energy, as was already observed and described for {\ddo} in Rieger \emph{et al.} \cite{Rieger2006}. With increasing $J$ and $K$ quantum numbers the spacing between the rotational energy levels increases, which reduces the Stark shifts.

\begin{figure*}
\includegraphics[width=.8\textwidth]{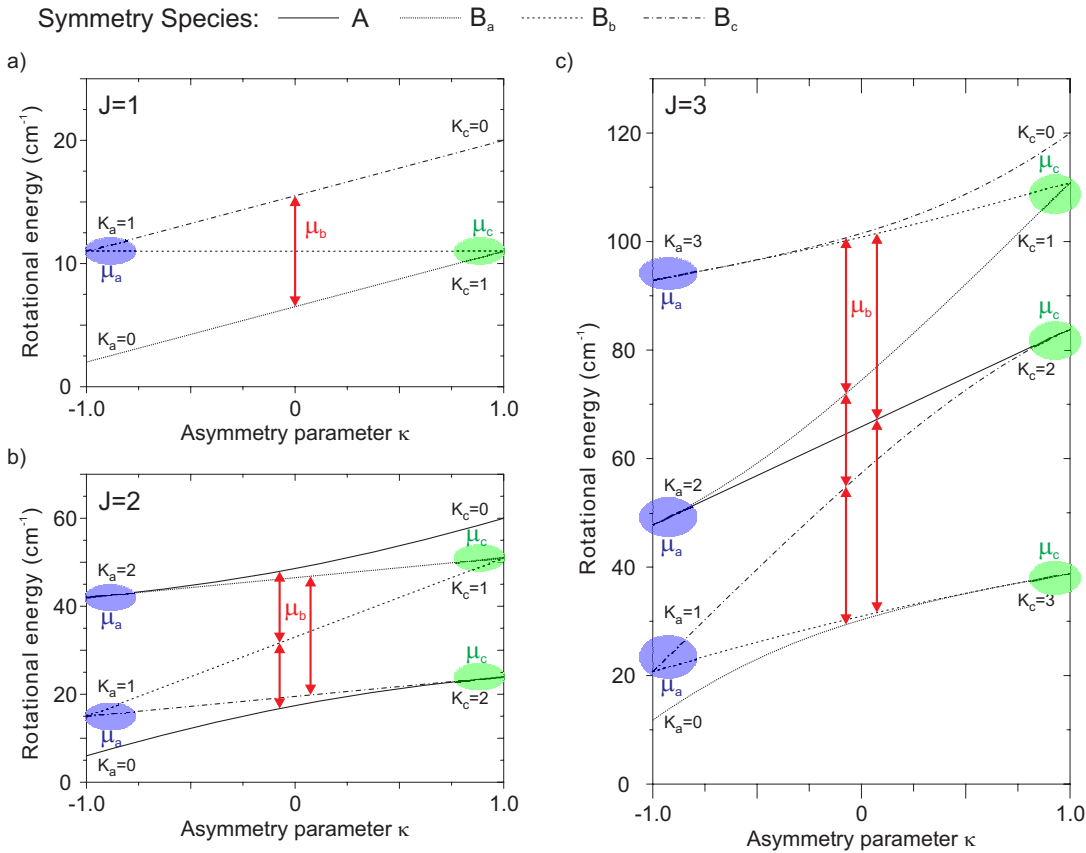}
\caption{\label{fig:EnergyLevelsAsymmTop}(Color online). Couplings due to the different dipole moment components in the asymmetric rotor. Shown are the energy levels of the asymmetric rotor as a function of the asymmetry parameter $\kappa$, for different $J$ quantum numbers. The symmetry of the rotational states, represented by the symmetry species of the four group $D_2$, is indicated by the dashing of the curves. A dipole moment $\mu_a$ along the $a$-axis mainly couples states which are degenerate in the prolate symmetric top, i.e.\ $\kappa=-1$, while a dipole moment $\mu_c$ along the $c$-axis mainly couples states degenerate in the oblate symmetric top, i.e.\ $\kappa=1$. This near degeneracy gives rise to linear Stark shifts for these states. In contrast, states which are coupled by a dipole moment $\mu_b$ along the $b$-axis, are always separated by an intermediate energy level, therefore never giving rise to a linear Stark shift. Rotational constants $A=10\,\mathrm{cm}^{-1}$ and $C=1\,\mathrm{cm}^{-1}$ were used, with $B=1\dots10\,\mathrm{cm}^{-1}$ varying linearly with $\kappa$.}
\end{figure*}

\begin{table}
\begin{ruledtabular}
\begin{tabular}{c|cc}
Dipole moment component &  \multicolumn{2}{c}{Coupled symmetry species}\\
\hline
$\mu_a$ &   $A \leftrightarrow B_a$ &   $B_b \leftrightarrow B_c$\\
$\mu_b$ &   $A \leftrightarrow B_b$ &   $B_a \leftrightarrow B_c$\\
$\mu_c$ &   $A \leftrightarrow B_c$ &   $B_a \leftrightarrow B_b$
\end{tabular}
\end{ruledtabular}
\caption{\label{tab:DipoleCoupling}Couplings between the different symmetry species of the asymmetric rotor belonging to the four group $D_2$ induced by the different dipole moment components, when an external electric field is applied along the $Z$-axis.}
\end{table}

More generally, the Stark shift properties discussed above can be understood for an arbitrary asymmetric top molecule by examining its rotational energy level structure, the symmetry properties of its energy levels, and the couplings brought about by the different components of the molecular electric dipole moment. These couplings between the different symmetry species are given in Table~\ref{tab:DipoleCoupling} for an external electric field along the $Z$-axis. Figure~\ref{fig:EnergyLevelsAsymmTop} shows the energy level structure of an asymmetric top molecule as a function of the asymmetry parameter $\kappa$ for fixed $J$ quantum numbers. The symmetry of the rotational energy levels, represented by the symmetry species $A, B_a, B_b, B_c$ of the four group $D_2$, is indicated. The couplings induced by the different dipole moment components $\mu_g$ according to Table~\ref{tab:DipoleCoupling} are also indicated in the figure.

As can be seen from Fig.~\ref{fig:EnergyLevelsAsymmTop}, for a near-prolate asymmetric top, $\kappa \approx -1$, the energy levels with same $K_a$ are near degenerate. These states are coupled by a dipole moment $\mu_a$ along the $a$-axis. This gives rise to linear Stark shifts, once the Stark interaction overcomes the splitting of these energy levels in the asymmetric rotor. Even in the most asymmetric case $\kappa =0$, the splitting between the states with highest rotational energy, $K_{a}=J,\,K_{c}=0,1$, decreases with increasing $J$, as can be seen by comparing the energy level curves for increasing values of the $J$ quantum number. This diminishing splitting gives rise to linear Stark shifts in the limit of large $J$ for a dipole moment $\mu_a$ along the $a$-axis, even in this most asymmetric case.

Similar arguments hold for a near-oblate asymmetric top, $\kappa \approx 1$. Here, the energy levels with same $K_c$ are near degenerate. These energy levels are coupled by a dipole moment $\mu_c$ along the $c$-axis, leading to linear Stark shifts as discussed for the near-prolate case. Once again, in the most asymmetric case $\kappa =0$ linear Stark shifts will occur in the limit of large $J$, however for different rotational states. The states $K_a=0,1,\,K_c=J$, being coupled by a dipole moment $\mu_c$ along the $c$-axis, show a decreasing splitting with increasing $J$, which can then give rise to linear Stark shifts.

The situation is, however, completely different for a dipole moment $\mu_b$ oriented along the $b$-axis. As can be seen from Fig.~\ref{fig:EnergyLevelsAsymmTop}, the energy levels which are coupled by such a dipole moment are always separated by another energy level in between. Therefore, the splitting between states coupled by a dipole moment along the $b$-axis never approaches zero within one $J$ system, independent of the value of the asymmetry parameter $\kappa$. More severely, the splitting between these energy levels coupled by a dipole moment $\mu_b$ along the $b$-axis even increases with increasing $J$. As can be seen from perturbation theory, this reduces the Stark shift for increasing rotational energy (large $J$). Of course, one could consider electric fields large enough to cause the Stark interaction to overcome this splitting of rotational states. Nonetheless, this generally does not lead to linear Stark shifts for low-field-seeking states. Coupling to other $J$ states becomes relevant first, forcing all states to become high-field-seeking.

The Stark shift properties examined in the preceding paragraphs can also be understood from a purely classical point of view. For a classical rotating body, the rotation is stable around the axis of least and largest moment of inertia, i.e.\ the $a$- and $c$-axis. Around the axis of intermediate moment of inertia no stable rotation is possible. If the dipole moment is oriented along the $a$- or $c$-axis, as is the case for a symmetric top or e.g.\ for an a-type asymmetric rotor such as formaldehyde (H$_2$CO), there exists a non-zero expectation value of the projection of the dipole moment on the space-fixed electric field axis, except for a classical motion with the axis of rotation perpendicular to the external field axis. This projection of the dipole moment can directly interact with the applied electric field, giving rise to a linear Stark shift. Conversely, if the dipole moment is oriented along the $b$-axis, i.e.\ the axis of intermediate moment of inertia, the expectation value of the projection of the dipole moment on the electric field axis will be zero. Hence, the external electric field first has to hinder the rotation and orient the dipole, which becomes more and more difficult with increasing rotational energy and angular momentum. Therefore, the magnitude of the oriented dipole decreases with increasing rotational energy, in accordance with the description given by quantum mechanics. This oriented dipole moment can then interact with the electric field. This second-order interaction gives rise to a quadratic Stark shift for $b$-type rotors. The same argument holds for linear heteronuclear molecules with a permanent electric dipole moment, where the rotation averages out the projection of the dipole moment. Hence, linear molecules exhibit a quadratic Stark effect \emph{unless} an electronic angular momentum parallel to the electric dipole moment $\vec{\mu}$ is present, as e.g.\ in $\Pi$ states of NH, OH or CO$^*$. However, note that in the case of accidental degeneracies also a $b$-type asymmetric rotor may show linear Stark shifts for certain states \footnote{For an example see, e.g., the Stark energy curves of the $b$-type asymmetric rotor CF$_2$H$_2$ in Fig.~2 in Hain \emph{et al.} \cite{Hain1999}. There, the state \ket{J=2,\tau=2,M=0} is near degenerate with \ket{J=1,\tau=0,M=0}, to which it is coupled by its dipole moment along the $b$-axis.}. The mechanical analogue of these accidental degeneracies, if existing, is not evident. Overall one can summarize that even molecules such as the three water isotopologs {\hho}, {\ddo} and {\hdo} which seem very similar at a first glance, can show surprisingly different behavior when exposed to an external electric field.

\section{Calculation of the guided flux}
\label{sec:calcGuide}

With the Stark shifts of the different isotopologs at hand we are now ready to calculate their guided fluxes. The total guided flux of a molecular species can be calculated by summing over the contributions to the flux of all individual internal states, weighted with their thermal occupation in the source. The thermal occupation of a rotational state \ket{J,\tau,M} in the source is given by
\begin{equation}
p_{J \tau M}=\frac{1}{Z}g_M g_I \exp{(-E_{J \tau M}/k_BT)},
\end{equation}
with the partition function
\begin{equation*}
Z=\sum_{J \tau M}g_M g_I \exp{(-E_{J \tau M}/k_BT)}.
\end{equation*}
$E_{J \tau M}$ is the rotational energy, $T$ the source temperature,  $g_M$ the $M$ degeneracy factor of the state, and $g_I$ the nuclear spin degeneracy factor. Nuclear spin degeneracy factors for the different isotopologs are listed in Table~\ref{tab:NuclSpinFactors}.
\begin{table}
\begin{ruledtabular}
\begin{tabular}{c|ccc}
$K_aK_c$                &{\hho}   &{\ddo}   &{\hdo}\\
\hline
$ee$, $oo$      &1        &6        &1\\
$eo$, $oe$      &3        &3        &1\\
\end{tabular}
\end{ruledtabular}
\caption{\label{tab:NuclSpinFactors}Nuclear spin degeneracy factors $g_I$ for the three water isotopologs {\hho}, {\ddo} and {\hdo} \cite{Townes:Microwave}. $e$ and $o$ refer to the parity (even or odd) of the quantum numbers $K_a$ and $K_c$ respectively.}
\end{table}

\begin{figure}
\includegraphics[width=1.\columnwidth]{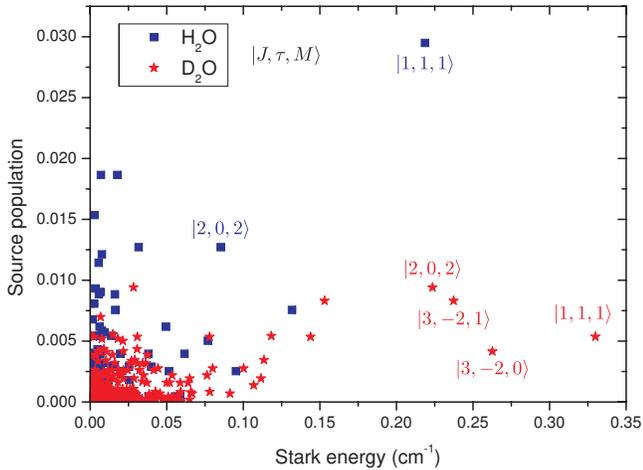}
\caption{\label{fig:CalcSourcePopStarkShift}(Color online). Calculated thermal populations in the source $p_{J\tau M}$ at a source temperature of 293\,K, and Stark shifts $\Delta W_{J\tau M}^s$ at an electric field of 130\,kV/cm for the rotational states \ket{J,\tau,M} of {\hho} and {\ddo}. States with large contributions to the guided flux are labeled with quantum numbers.}
\end{figure}

The guided flux for a given guiding electric field can now be readily expressed as
\begin{equation}
\Phi = N_0 \sum_{J \tau M} p_{J \tau M} f(\Delta W_{J \tau M}),
\end{equation}
with the guidable fraction of a rotational state \ket{J \tau M} given by $f(\Delta W_{J \tau M}^s) \propto (\Delta W_{J \tau M}^s)^2$ as shown in Section~\ref{sec:velfilter}, and $N_0$ the number of molecules injected into the guide per second. Figure~\ref{fig:CalcSourcePopStarkShift} compares Stark shifts and calculated source populations of individual rotational states for {\hho} and {\ddo}. From this figure it can already be anticipated that in the case of {\hho} the state \ket{J=1,\tau=1,M=1} will contribute strongly to the guided flux. The reason for the large thermal population of this state as compared to the \ket{J=1,\tau=1,M=1} state in {\ddo} can be found in the different nuclear spin statistics of the two isotopologs. For such considerations based on symmetry properties it is advantageous to transform from \ket{J,\tau,M} to the quantum numbers \ket{J,K_a,K_c,M}. For the state \ket{J=1,\tau=1,M=1} this results in the corresponding quantum numbers \ket{J=1,K_a=1,K_c=0,M=1}. According to Table~\ref{tab:NuclSpinFactors}, this state is favored in {\hho} by $3:1$, while it is unfavored in {\ddo} by $3:6$.

\begin{figure}
\includegraphics[width=1\columnwidth]{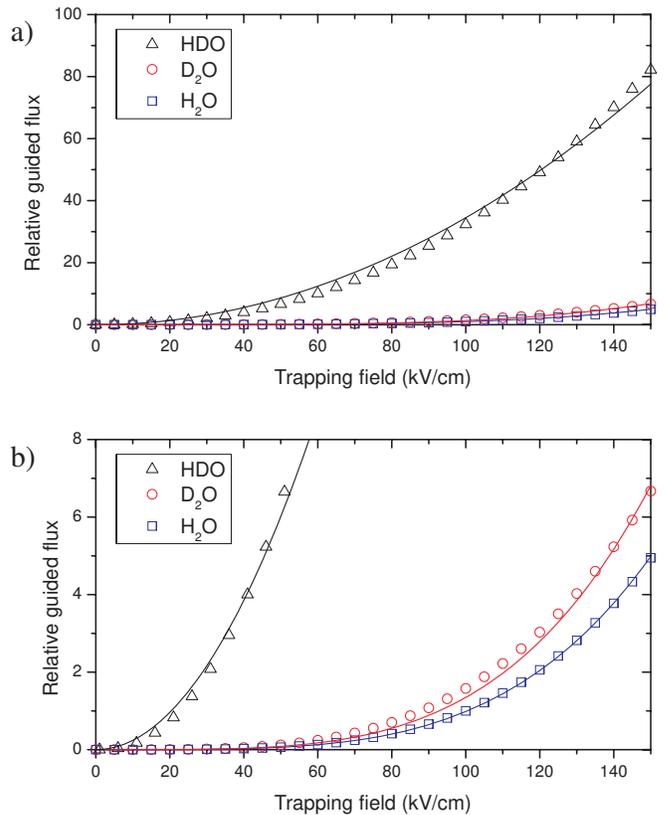}
\caption{\label{fig:CalcGuidedFlux}(Color online). a) Calculated guided fluxes of the different water isotopologs as a function of the trapping electric field. The symbols are the results of the calculation described in Section \ref{sec:calcGuide}, whereas the solid curves are quartic and quadratic fits to the calculated flux. Guided fluxes are normalized to the flux of {\hho} at a trapping field of 100\,kV/cm. The guided flux of {\hdo} shows a quadratic dependence on the trapping electric field, whereas for {\hho} and {\ddo} a quartic dependence is found. b) Zoom-in to compare the calculations of the guided flux for {\hho} and {\ddo} to a quartic curve.}
\end{figure}

The expected guided fluxes of the different isotopologs calculated this way are depicted in Fig.~\ref{fig:CalcGuidedFlux} as a function of the electric field. The guided fluxes of the different isotopologs differ remarkably. Furthermore, not only the flux but also the electric field dependence is clearly different. While {\hho} and {\ddo} show a quartic dependence of the guided flux, caused by quadratic Stark shifts, the electric field dependence of {\hdo} shows a mainly quadratic behavior, indicating the presence of linear Stark shifts.

\begin{table}
{\hho}
\vspace{1em}
\begin{ruledtabular}
\begin{tabular}{ccc|dddd}
$J$ &   $\tau$  &   $M$ &
\multicolumn{1}{c}{\minitab[c]{$E_{rot}$\\(cm$^{-1}$)}}&
\multicolumn{1}{c}{\minitab[c]{$p_S$ \\ (\%)}}    &
\multicolumn{1}{c}{\minitab[c]{$\Delta W^s$\\(cm$^{-1}$)}}   &
\multicolumn{1}{c}{\minitab[c]{$p_G$ \\ (\%)}} \\
\hline
1   &1  &1  & 42.3   &2.9    &0.13   &79.3\\
2   &0  &2  & 95.7   &0.8    &0.08   &7.4\\
3   &1  &3  & 213.6  &1.3    &0.05   &5.1\\
\end{tabular}
\end{ruledtabular}
\vspace{1em}
{\ddo}
\vspace{1em}
\begin{ruledtabular}
\begin{tabular}{ccc|dddd}
$J$ &   $\tau$  &   $M$ &
\multicolumn{1}{c}{\minitab[c]{$E_{rot}$\\(cm$^{-1}$)}}&
\multicolumn{1}{c}{\minitab[c]{$p_S$ \\ (\%)}}    &
\multicolumn{1}{c}{\minitab[c]{$\Delta W^s$\\(cm$^{-1}$)}}   &
\multicolumn{1}{c}{\minitab[c]{$p_G$ \\ (\%)}} \\
\hline
1   &1  &1  & 22.7  &0.54   &0.20   &21.3\\
3   &-2 &1  & 74.5  &0.83   &0.16   &21.2\\
2   &0  &2  & 49.3  &0.94   &0.13   &17.1\\
3   &-2 &0  & 74.5  &0.42   &0.18   &13.2\\
3   &-2 &2  & 74.5  &0.83   &0.10   &8.0\\
\end{tabular}
\end{ruledtabular}
\vspace{1em}
{\hdo}
\vspace{1em}
\begin{ruledtabular}
\begin{tabular}{ccc|dddd}
$J$ &   $\tau$  &   $M$ &
\multicolumn{1}{c}{\minitab[c]{$E_{rot}$\\(cm$^{-1}$)}}&
\multicolumn{1}{c}{\minitab[c]{$p_S$ \\ (\%)}}    &
\multicolumn{1}{c}{\minitab[c]{$\Delta W^s$\\(cm$^{-1}$)}}   &
\multicolumn{1}{c}{\minitab[c]{$p_G$ \\ (\%)}} \\
\hline
3   &3   &3  & 234.9  &0.45   &0.83   &15.0\\
2   &2   &2  & 109.8  &0.83   &0.59   &14.3\\
4   &4   &4  & 407.0  &0.19   &0.89   &7.5\\
3   &3   &2  & 234.9  &0.45   &0.55   &6.6\\
4   &2   &4  & 297.8  &0.33   &0.58   &5.5\\
\end{tabular}
\end{ruledtabular}

\caption{\label{tab:PopStatesGuide}Properties of selected rotational states (population in the guide$>$5\,\%) of the three water isotopologs {\hho}, {\ddo} and {\hdo}\@. $E_{rot}$: zero field rotational energy, $p_S$: Source population for a temperature of 293\,K, $\Delta W^s$: Stark shift at an electric field of 100\,kV/cm, $p_G$: population in the electric guide at an applied electric field of 100\,kV/cm.}
\end{table}

The calculations of the guided fluxes allow us to deduce the populations of individual rotational states in the guided beam. In a recent experiment these calculated state populations were experimentally verified by collinear ultraviolet spectroscopy in a cold guided beam of formaldehyde H$_2$CO \cite{Motsch2007}. For the water isotopologs, populations of individual rotational states with contributions larger than 5\% of the total flux of the considered isotopolog are listed in Table~\ref{tab:PopStatesGuide}. Remarkably, the single state \ket{J=1,\tau=1,M=1} of {\hho} contributes $\approx$80\,\% to the guided flux using a room temperature source. Similarly, in {\ddo} the most populated state \ket{J=1,\tau=1,M=1} contributes $\approx$21\,\% to the total flux. The reason for these large populations as compared to a molecule such as formaldehyde (H$_2$CO) \cite{Motsch2007} with similar size rotational constants but mainly linear Stark shifts can be found in the quadratic Stark shift behavior. As can be seen from Fig.~\ref{fig:StarkShifts}, the size of the Stark shift decreases with rotational energy for a molecule with a quadratic Stark shift, leaving only few low-energy states with largest Stark shifts for guiding \cite{Rieger2006}.

\section{Experimental Setup}
\label{sec:exp}
\begin{figure}
\includegraphics[width=1\columnwidth]{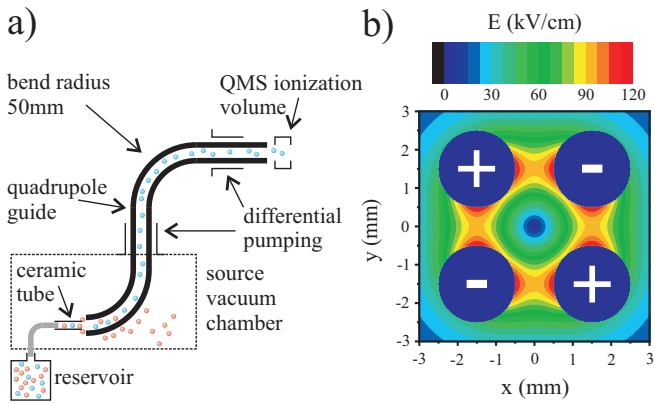}
\caption{\label{fig:ExpSetup}(Color online). Experimental setup. a) Water molecules from the thermal reservoir enter the electric guide. Slow molecules are trapped in the quadrupole field and guided to an ultra-high-vacuum chamber, where they are detected by the quadrupole mass spectrometer. b) Electric field distribution in the quadrupole guide for $\pm$5\,kV electrode voltage, resulting in a trapping electric field exceeding 93\,kV/cm.}
\end{figure}
To compare with theory, experiments with the three water isotopologs {\hho}, {\hdo\ and {\ddo} were performed. As shown in Fig.~\ref{fig:ExpSetup}, a setup similar to the one described previously \cite{Junglen2004a,Motsch2007} is used for the experiment. Water molecules are injected into the quadrupole guide through a ceramic tube with diameter 1.5\,mm and a length of 9.5\,mm. The pressure in the reservoir is kept at a fixed value of 0.10\,mbar via a stabilized flow valve, resulting in a gas flow of $1\times 10^{-4}$\,mbar$\cdot$l/s. Since water has a sufficiently high vapor pressure of $\approx25$\,mbar at room temperature \cite{HandbookChemPhys71st}, no heating of its container is necessary. The constituents of the gas injected through the tube can be monitored by a residual gas analyzer placed in the source vacuum chamber. The guide is composed of four stainless steel electrodes of 2\,mm diameter separated by a distance of 1\,mm. For an electrode voltage of $\pm$5\,kV, a trapping electric field of around 93\,kV/cm is generated. Transversely slow molecules in a lfs state are trapped by the enclosing high electric fields. A longitudinal cut-off velocity is obtained by bending the guide. Slow molecules are guided around two bends with a radius of curvature of 5\,cm and through two differential pumping stages to an ultrahigh-vacuum chamber, where they are detected by a quadrupole mass spectrometer (QMS). The molecules are ionized by electron impact in a cross-beam ion source and mass selected in the analyzer. In the final stage of the QMS single ion counting is performed.

Measurements of {\ddo} were performed with isotopically pure {\ddo}. For measurements of {\hdo}, mixtures of {\hho} and {\ddo} with a ratio of 1:1  and 4:1 were used. These result in a {\hdo} fraction of $\approx$48\,\% and $\approx$27\,\%, respectively, coming from the source as measured with the residual gas analyzer. When using these mixtures in the source, all three isotopologs are guided simultaneously. However, due to its much larger Stark shift for some rotational states (see Section~\ref{sec:calcStark}, and Fig.~\ref{fig:StarkShifts}), {\hdo} is preferentially guided. This measurement with the mixture of different isotopologs allows to extract contributions of {\hdo} and {\ddo}. The contribution of {\hho} is also visible, although it is shadowed by fragments of {\ddo} and {\hdo} in the QMS at mass 18, 17 and 16 amu where {\hho} is detected. Measurements of {\hho} were performed with pure {\hho} to avoid these unwanted contributions of {\ddo} and {\hdo}.

\section{Results and Discussion}
\label{sec:results}

\begin{figure}
\includegraphics[width=1\columnwidth]{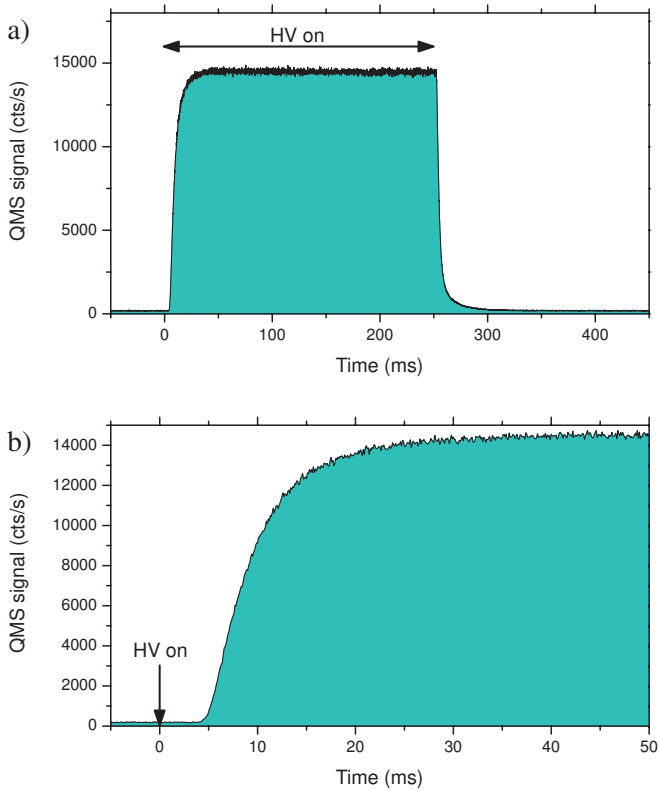}
\caption{\label{fig:toftrace}(Color online). a) Time of flight trace measured with {\hdo} for an electrode voltage of $\pm$5\,kV. High voltage (HV) is switched on for an interval of 250\,ms, with a 50\,\% duty cycle. The QMS signal starts rising with a time delay after switching on HV, until a steady state is reached. This time delay corresponds to the fastest molecules' travel time from the nozzle to the QMS. The steady-state value is used for voltage dependence measurements. The slow decay after switching off the high voltage is due to the fact that the last section of the guide closest to the QMS is not switched to avoid electronic pick up on the QMS. Hence, molecules in this part can arrive at the QMS even after the HV has been switched off in the first section. b) Zoom into the rising slope of the signal, from which a velocity distribution can be derived. These plots show raw data, i.e.\ no background subtraction has been applied, to illustrate the excellent signal to noise ratio.}
\end{figure}

The experiments are performed as a series of time of flight measurements. Here, the high voltage (HV) is switched on and off repeatedly in a fixed timing sequence. A typical time of flight trace is shown in Fig.~\ref{fig:toftrace}. To subtract background contributions, the guided signal for the different isotopologs is determined from the difference in the steady state QMS signal with HV applied to the guide and HV switched off.

Before comparing the signals of guided molecules to the predictions of the theory presented in Section~\ref{sec:calcGuide}, the detection process is considered in some more detail. The QMS used for detection of the guided molecules operates as a residual gas analyzer, which measures the densities of individual constituents in the recipient. The signals of guided molecules observed in previous \cite{Rangwala2003,Junglen2004a} as well as in the present experiments are, however, fully compatible with the predicted electrode voltage dependence of the guided flux.

The solution to this surprising fact lies in the pressure dependence. The model of velocity filtering presented in Section~\ref{sec:velfilter} and \ref{sec:calcGuide} assumes that the molecules from a thermal ensemble are transferred into the guide while preserving their velocity distributions. Measurements performed with deuterated ammonia ND$_3$ for very small inlet pressures indeed show the characteristic voltage dependence expected for a density measurement \cite{Motsch2008c}. For increasing inlet pressure, however, the voltage dependence changes, and the signal more and more resembles a flux measurement. Here, collisions of molecules in the nozzle and in the "high pressure" region directly behind the nozzle become more likely. Slow molecules are more likely removed from the ensemble, effectively leading to a "boosting" of the beam. This boosting is also visible in the velocity distributions of the guided beam presented in Section~\ref{subsec:results:veldist}.

From the pressure dependence observed in these measurements with ND$_3$ the following conclusion can be drawn: by including this boosting effect in the model, the voltage dependencies over the entire pressure range studied agree with a density measurement. For the conditions chosen in previous experiments as well as in the experiments described in this paper the electrode voltage dependence of the QMS signal resembles a flux measurement of a gas without boosting. Therefore, the outcome of the experiment can be directly compared to the theory for an ideal effusive source presented in Section~\ref{sec:velfilter} and \ref{sec:calcGuide}. To avoid any ambiguity, we refer to the signal of guided molecules as "molecule signal" or "detector signal".

To extract the relative molecule signals of the different isotopologs from the individual measurements, some corrections are necessary. First of all, the fragmentation of the guided molecules in the QMS has to be taken into account, since the measurement is performed at only one mass (18\,amu for {\hho}, 20\,amu for {\ddo}, and 19\,amu for {\hdo}). The correction is determined from the ion count rates of the different fragmentation products for pure {\hho} (18, 17, 16\,amu), pure {\ddo} (20, 18, 16\,amu), and for the mixture with high (48\,\%) {\hdo} content (19, 18, 17, 16\,amu), where {\hdo} dominates the molecule signal. Secondly, the ion count rate for each isotopolog is corrected for the relative contributions of the isotopologs injected into the guide as monitored by the residual gas analyzer in the source chamber. A velocity distribution of the guided molecules is determined from the rising slope of the time of flight trace.

\subsection{Electrode Voltage Dependency}

\begin{figure}
\includegraphics[width=1\columnwidth]{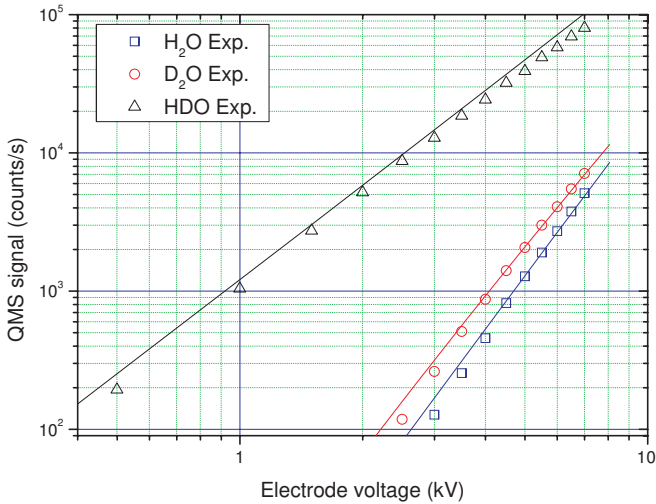}
\caption{\label{fig:GuidedFluxLog}(Color online).
Guided signal of {\hho}, {\ddo} and {\hdo} as a function of the applied electrode voltage. Shown as solid curves are the theoretically predicted signal dependencies, adjusted to fit {\ddo} data at $\pm$6\,kV electrode voltage. Note that only one global scaling factor is used for all theory curves. The different slopes of the curves in the double logarithmic plot directly indicate the different Stark shift properties. {\hho} and {\ddo} show a quartic dependence of guided signal on the applied electric field, indicating quadratic Stark shifts, whereas for {\hdo} a mainly quadratic dependence indicating linear Stark shifts is found.}
\end{figure}

As shown in Fig.~\ref{fig:GuidedFluxLog}, the electrode voltage dependence of the molecule signals for {\hho} and {\ddo} are well described by the calculations over more than two orders of magnitude. The calculations are scaled by only one global scaling factor to account for detection efficiencies and the amount of gas injected through the nozzle into the guide. Note that \emph{no relative scaling factor} between the calculations for {\hho}, {\ddo} and {\hdo} has been employed. This excellent agreement between experiment and theory verifies that the filtering process is well described by the model presented in Sections~\ref{sec:velfilter}, \ref{sec:calcStark} and \ref{sec:calcGuide}.

As can be seen from Fig.~\ref{fig:GuidedFluxLog}, {\hdo} shows the largest molecule signal of all three water isotopologs, as predicted from calculations. We experimentally determine a ratio of detector signals between {\hdo} and {\ddo} at a guiding field of 130 (100)\,kV/cm of 16.4 (18.9), where the calculations predict 14.7 (20.5). This is well within the overall uncertainty given e.g.\ by the determination of the relative contributions of isotopologs injected through the nozzle as derived from the residual gas analyzer signal. The measurements done with different {\hdo} amounts give count rates agreeing to within 5--10\,\%, which allows an estimate for the uncertainty of the residual gas analyzer corrections applied. Furthermore, the guided flux of {\hdo} largely stems from rotational states at higher rotational energies. These are therefore more strongly affected by centrifugal distortion corrections not included in the rigid rotor approximation being used, leading to energy level shifts and hence changes of Stark shifts which can affect the accuracy of the calculations.

The measured signals of guided {\hho} and {\ddo} show a quartic dependence on the applied electrode voltage, as was shown for {\ddo} already in a previous experiment \cite{Rieger2006}. The electrode voltage dependence for the molecule signal of {\hdo} is best described by a quadratic behavior, confirming the main contributions from states with linear Stark shifts predicted by calculations. This different dependence of the guided signal on the applied electric field is directly evident from different slopes in the double logarithmic plot shown in Fig.~\ref{fig:GuidedFluxLog}. Also the good agreement between calculations and the experiment over a wide range of applied electrode voltages resulting in changes of several orders of magnitude for the molecule signal is remarkable.

The detector signals of the water isotopologs can also be roughly compared to other molecules used so far. For deuterated ammonia ND$_3$, count rates of the order of $3\times10^5$\,cts/s were observed in the same setup at an electrode voltage of $\pm$5\,kV, using a reservoir at room temperature. This is to be compared to a guided signal of $\approx4\times10^4$\,cts/s for {\hdo}\@. The reason for this smaller flux can be found in the fact that for {\hdo} only few states exhibit linear Stark shifts and hence contribute to the guided flux, whereas ND$_3$ exhibits mainly linear Stark shifts. Furthermore the Stark shifts of {\hdo} are smaller due to the smaller dipole moment component along the $a$-axis as compared to the dipole moment of ND$_3$. Nevertheless, the fact that molecules with so different Stark shift properties can be efficiently filtered out of a thermal gas illustrates the wide applicability of the velocity filtering technique. For chemically stable polar molecules which can be supplied to the nozzle via the teflon tube, it is sufficient to bring them into the gas phase in a reservoir at a pressure in the 0.1--1\,mbar range. At smaller pressures guiding is still possible, however at the expense of reduced signals. Chemically very reactive molecules such as molecular radicals (typical examples are OH and NH) are in principle also guidable. Beams of molecular radicals can be loaded into a helium buffer gas cell, where they thermalize by collisions with the cryogenic helium gas \cite{Campbell2007}. The translationally and internally cooled down molecules can then be extracted by the electrostatic quadrupole guide \cite{vanBuuren2008} and made available for further experiments.

\subsection{Velocity Distributions}
\label{subsec:results:veldist}

\begin{figure}
\includegraphics[width=1\columnwidth]{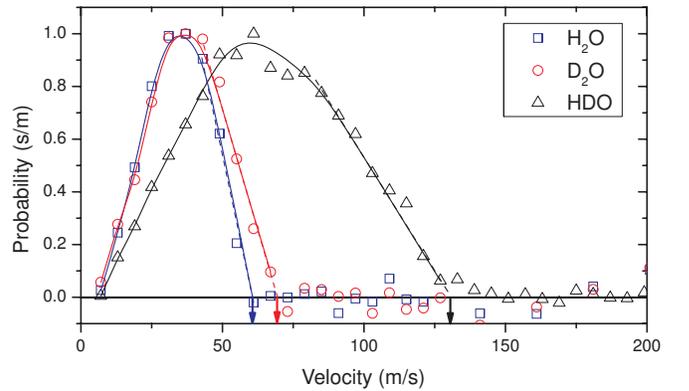}
\caption{\label{fig:VelDistAll}(Color online). Normalized velocity distributions for the water isotopologs {\hho}, {\ddo} and {\hdo} at an electrode voltage of $\pm$5\,kV\@. The velocity distributions are derived from the rising slope of time of flight measurements. The solid curves are a guide to the eye. Vertical arrows indicate the cut-off velocities obtained by linear extrapolation (dashed lines) of the high velocity side.}
\end{figure}

The different Stark shift properties of the various isotopologs are also evident from their velocity distributions. The cut-off velocity, i.e.\ the velocity of the fastest molecules which can still be guided, is given by the molecules' Stark shift. Hence, it depends on the rotational state of the molecule. Since for molecules from a thermal source many states contribute to the guided flux, the cut-off velocity is given by the state with the largest Stark shift. In the experiment the cut-off velocity is determined from a linear extrapolation of the high-velocity side of the velocity distribution towards zero. From the measurements performed at an electrode voltage of $\pm$5\,kV it can be seen in Fig.~\ref{fig:VelDistAll} that the maximum of the distribution as well as the cut-off velocity shifts towards higher velocities from {\hho} and {\ddo} to {\hdo}\@. This is caused by the much larger Stark shifts of {\hdo} as compared to {\hho} and {\ddo} (see Fig.~\ref{fig:StarkShifts}). The wide velocity distribution for {\hdo} is caused by the large Stark shifts found for this isotopolog and by the used bend radius of 5\,cm. We experimentally determine cut-off velocities of 60\,m/s for {\hho}, 69\,m/s for {\ddo}, and 130\,m/s for {\hdo} at an electrode voltage of $\pm5\,\mathrm{kV}$. Calculated Stark shifts of states contributing to the guided flux (similar to Table~\ref{tab:PopStatesGuide}, but for a guiding field of 93\,kV/cm reached at $\pm$5\,kV electrode voltage) result in cut-off velocities of 57\,m/s for {\hho}, 67\,m/s for {\ddo} and 125\,m/s for {\hdo}, which is in good agreement with measurement. For {\hdo} states with large Stark shifts of up to 0.80\,cm$^{-1}$ at $\pm5\,\mathrm{kV}$ electrode voltage are predicted to contribute with $\approx 20\%$ to the guided flux. These states might be responsible for the small but non-zero signal in the {\hdo} velocity distribution between 130\,m/s and 150\,m/s, being supported by the fact that a cut-off velocity of 150\,m/s corresponds to a Stark shift of 0.80\,cm$^{-1}$. Note that this signal exceeding the linear extrapolation to the falling slope of the velocity distribution is present in the velocity distributions of {\hdo} measured at $\pm3$\,kV, $\pm5$\,kV and $\pm7$\,kV\@.

On the low velocity side of the velocity distribution a linear extrapolation does not cut the horizontal axis at zero velocity. We attribute this to collisions of the molecules in the vicinity of the effusive source, which removes the slowest molecules from the thermal distribution, effectively leading to a "boosting" of the molecules entering the guide. This is supported by measurements with other gases such as ND$_3$, where an influence of the reservoir pressure on the number of molecules with low velocities was observed.

\begin{figure}
\includegraphics[width=1\columnwidth]{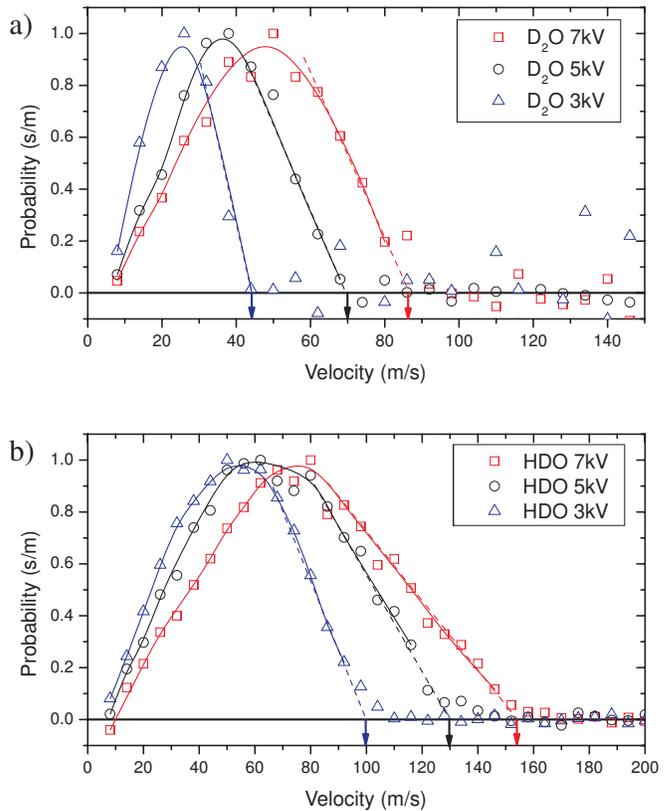}
\caption{\label{fig:VelDistVoltDep}(Color online). Normalized velocity distributions for {\ddo} and {\hdo} as a function of electrode voltage ($\pm3$\,kV, $\pm5$\,kV, $\pm7$\,kV). The solid curves are a guide to the eye. The vertical arrows indicate the cut-off velocity, which is determined from a linear approximation (dashed lines) to the high velocity side.}
\end{figure}

\begin{figure}
\includegraphics[width=1\columnwidth]{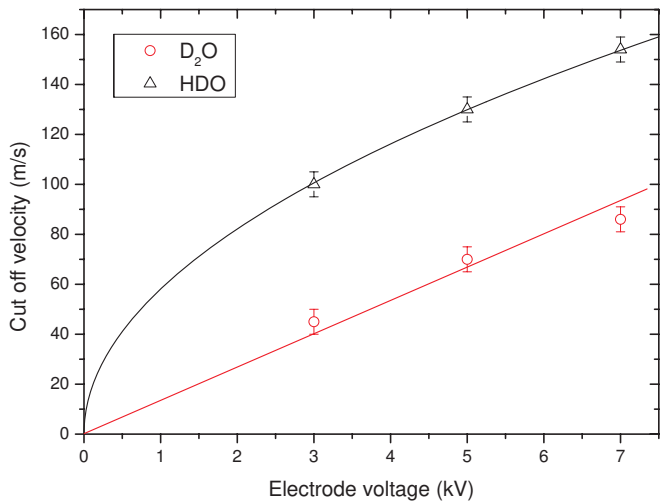}
\caption{\label{fig:VelDistCutOff}(Color online). Voltage dependence of cut-off velocities for {\ddo} and {\hdo}\@. For {\ddo} the line is a linear fit, for {\hdo} the solid curve is a fit of a square root dependence. These dependencies are expected for quadratic ({\ddo}) and linear ({\hdo}) Stark shifts respectively.}
\end{figure}

Measurements of velocity distributions for different electrode voltages ($\pm$3\,kV, $\pm$5\,kV, $\pm$7\,kV) were performed for {\ddo} and {\hdo}, and are shown in Fig.~\ref{fig:VelDistVoltDep}. In the data a shift of the maximum of the distribution and of the cut-off velocity, i.e.\ the maximum velocity of molecules which can still be guided, can be seen. This is to be expected, since larger voltages applied to the guide electrodes result in a deeper trapping potential, hence faster molecules are guided. Regarding the dependence of the cut-off velocities on the applied electrode voltage in Fig.~\ref{fig:VelDistCutOff}, a linear dependence in the case of {\ddo} and a square-root dependence in the case of {\hdo} is observed. As was shown in Section \ref{sec:velfilter}, the cut-off velocity depends on the square-root of the Stark shift at the applied guiding field. Hence, a linear dependence as for {\ddo} is found for molecules with quadratic Stark shift, while the square-root dependence is found for molecules as {\hdo} where states with linear Stark shifts dominate the guided flux.

\section{Summary and Outlook}
\label{sec:outlook}

\begin{table}
\begin{ruledtabular}
\begin{tabular}{c|ccc}
Representation & \Roem{1}$^r$ & \Roem{2}$^r$ & \Roem{3}$^r$ \\
\hline
x & b & c & a\\
y & c & a & b\\
z & a & b & c\\
\end{tabular}
\end{ruledtabular}
\caption{The three possible connections between the principal axes of the molecular moment of inertia tensor ($a$, $b$, $c$) and the right-handed body-fixed coordinate system ($x$, $y$, $z$). The $^r$ denotes the choice of right-handed coordinate systems (adapted from \cite{King1943}).}
\label{tab:Representations}
\end{table}

\begin{table}
\begin{ruledtabular}
\begin{tabular}{c|ccc}
& \multicolumn{3}{c}{Representation}\\
& \Roem{1}$^r$ & \Roem{2}$^r$ &\Roem{3}$^r$ \\
\hline
$F$ & $\frac{1}{2} (\kappa -1)$ & 0 &$\frac{1}{2} (\kappa +1)$\\
$G$ & 1 & $\kappa$ & -1\\
$H$ & $-\frac{1}{2} (\kappa +1)$ & 1 & $\frac{1}{2} (\kappa -1)$\\
\end{tabular}
\end{ruledtabular}
\caption{Coefficients used in the matrix elements of the reduced Hamiltonian (Eq.~\ref{eq:AsymmRotRedHamiltonianMatrixElements}) for the different representations listed in Table~\ref{tab:Representations} (adapted from \cite{King1943}).}
\label{tab:FGH}
\end{table}

To summarize, we have produced continuous cold guided beams of the water isotopologs {\hho}, {\ddo} and {\hdo} by electrostatic velocity filtering. We discuss in detail the influence of molecular parameters such as rotational constants and orientations of electric dipole moments on the behavior of the molecule in an external electric field. Based on this discussion of molecular Stark shifts, we develop the theory for the electrostatic velocity filtering process. We find excellent agreement between the experimentally observed signals and our calculations of guided fluxes, confirming that the filtering process is well described by the presented model. The molecule signals of the different isotopologs show quartic and quadratic electrode voltage dependencies respectively caused by quadratic Stark shifts for {\hho} and {\ddo} and by linear Stark shifts of the states contributing most to the guided flux in {\hdo}\@. These different Stark shift properties can also be seen from the dependence of the cut-off velocity on the applied guiding electric field as measured for {\ddo} and {\hdo}\@. Furthermore, calculations of guided fluxes allow to deduce populations of individual rotational states in the guided beam. These can be as high as 80\,\% in the case of {\hho} for a room-temperature source. Overall, this shows that velocity filtering by an electric guide is a technique applicable to a wide variety of molecular species.

In the presented experiment, the internal state distribution was only inferred from calculations based on the good agreement between theory and experiment. However, internal state diagnostics of cold guided water beams or any other guided species should be possible by transferring the depletion spectroscopy technique used for formaldehyde in the near-ultraviolet \cite{Motsch2007} to either the infrared using vibrational transitions or to the microwave domain using purely rotational transitions. Such additional state-dependent detection will be beneficial for studies of, e.g., collision processes with or between cold molecules. Furthermore, with cold water beams at hand it should be possible to investigate in the laboratory processes such as, e.g., ice formation on dust grains under conditions present in the interstellar medium.

\begin{acknowledgments}
Support by the Deutsche Forschungsgemeinschaft through the excellence cluster "Munich Centre for Advanced Photonics" and EuroQUAM (Cavity-Mediated Molecular Cooling) is acknowledged.
\end{acknowledgments}

\appendix*
\section{Tables}
\label{sec:AppendixTables}

In this appendix we provide tables \ref{tab:Representations}--\ref{tab:PhiFg}.

\begin{widetext}
\phantom{just some space}

\begin{table*}[b]
\begin{ruledtabular}
\begin{tabular}{c|ccc}
& \multicolumn{3}{c}{$ J'$}\\
& $ J+1$ & $ J$ & $ J-1$\\
\hline
$\bra{J}\Phi_{Fg}\ket{J'}$ & $\left[4\left(J+1\right)\left[\left(2J+1\right)\left(2J+3\right)\right]^{1/2}\right]^{-1}$&
$\left[4J\left(J+1\right)\right]^{-1}$&
$\left[4J\left(4J^2-1\right)^{1/2}\right]^{-1}$\\

$\bra{JK}\Phi_{Fz}\ket{J'K}$ &
$2 \left[\left(J+1\right)^2-K^2\right]^{1/2}$&
$2K$ &
$-2\left[J^2-K^2\right]^{1/2}$
\\

$\bra{JK}\Phi_{Fy}\ket{J',K\pm1}$ &
\multirow{2}{*}
{$\mp \left[\left(J\pm K +1\right)\left(J\pm K +2\right)\right]^{1/2} $}&
\multirow{2}{*}
{$\left[\left(J\mp K\right)\left(J\pm K +1\right)\right]^{1/2} $} &
\multirow{2}{*}
{$\mp \left[\left(J\mp K\right)\left(J\mp K -1\right)\right]^{1/2} $}\\
$=\mp i\bra{JK}\Phi_{Fx}\ket{J',K\pm1}$ & & & \\

$\bra{JM}\Phi_{Zg}\ket{J'M}$&
$2\left[(J+1)^2-M^2\right]^{1/2}$&
$2M$&
$-2\left[J^2-M^2\right]^{1/2}$
\\

$\bra{JM}\Phi_{Yg}\ket{J',M\pm1}$&
\multirow{2}{*}
{$\mp \left[\left(J\pm M +1\right)\left(J\pm M +2\right)\right]^{1/2}$}&
\multirow{2}{*}
{$\left[\left(J\mp M\right)\left(J\pm M +1\right)\right]^{1/2} $}&
\multirow{2}{*}
{$\mp \left[\left(J\mp M\right)\left(J\mp M -2\right)\right]^{1/2} $}\\
$=\pm i\bra{JM}\Phi_{Xg}\ket{J',M\pm1}$ & & &\\
\end{tabular}
\end{ruledtabular}
\caption{Direction cosine matrix elements $\bra{J,K,M}\Phi_{Fg}\ket{J'K'M'}$ for the symmetric rotor (adapted from \cite{Cross1944}).}
\label{tab:PhiFg}
\end{table*}

\end{widetext}

\end{document}